\documentclass[%
 aip,
 jmp,%
 amsmath,amssymb,
reprint,%
]{revtex4-1}

\usepackage{graphicx}% Include figure files
\usepackage{dcolumn}% Align table columns on decimal point
\usepackage{bm}% bold math
\usepackage[usenames]{color}
\definecolor{cRef1}{rgb}{1.0,0.0,0.0}
\usepackage{booktabs}
\usepackage{ctable}
\usepackage{multirow}
\def\Wcmcm{\mbox{\rm Wcm$^{-2}$}}
\def\Wcmcmmu{\mbox{\rm Wcm$^{-2}\mu$m$^2$}}
\begin{document}

%\preprint{AIP/123-QED}

\title{Collisionless absorption, hot electron generation, 
and energy scaling in intense laser-target interaction}
% Force line breaks with \\
%\thanks{Footnote to title of article.}

\author{T. Liseykina}
\email{tatyana.tiseykina@uni-rostock.de}
\affiliation{Institut f\"{u}r Physik, Universit\"{a}t Rostock, Universit\"{a}tsplatz 3, 18051 Rostock, Germany}
\affiliation{Institute of Computational Technologies SD RAS, Acad. Lavrentjev ave. 6, 630090 Novosibirsk, Russia}

\author{P. Mulser}%
\affiliation{Theoretical Quantum Electronics, Technische Universit\"{a}t Darmstadt, 64289 Darmstadt,~Germany}
\author{M. Murakami}%
\affiliation{Institute of Laser Engineering, Osaka University, Osaka 565-0871, Japan}%

\begin{abstract}
Among the various attempts to understand collisionless absorption of intense and superintense ultrashort laser 
pulses a whole variety of models and hypotheses has been invented to describe the laser beam target interaction. 
In terms of basic physics collisionless absorption is understood now as the interplay of the oscillating laser 
field with the space charge field produced by it in the plasma. A first approach to this idea is realized in 
Brunel's model the essence of which consists in the formation of an oscillating charge cloud in the \nobreak{vacuum} 
in front of the target, therefore frequently addressed by the vague term "vacuum heating". The investigation 
of statistical ensembles of orbits shows that the absorption process is localized at the ion-vacuum interface 
and in the skin layer: Single electrons enter into resonance with the laser field thereby undergoing a phase 
shift which causes orbit crossing and braking of Brunel's laminar flow. This anharmonic resonance acts like 
an attractor for the electrons and leads to the formation of a Maxwellian tail in the electron energy spectrum. %140
Most remarkable results of our investigations are the Brunel like spectral hot electron distribution at the 
relativistic threshold, the minimum of absorption at $I\lambda^2 \cong (0.3-1.2)\times 10^{21}$ \Wcmcmmu\,
in the plasma target with the electron density of $n_e \lambda^2\sim 10^{23}$cm$^{-3}\mu$m$^2,$ the drastic 
reduction of the number of hot electrons in this domain and their reappearance in the highly relativistic 
domain, and strong coupling, beyond expectation, of the fast electron jets with the return current through 
Cherenkov emission of plasmons. The hot electron energy scaling shows a strong dependence on intensity in the 
moderately relativistic domain $I\lambda^2 \cong (10^{18} - 10^{20})$ \Wcmcmmu, a scaling in vague 
accordance with current published estimates in the range $I\lambda^2 \cong (0.14-3.5)\times 10^{21}$ \Wcmcmmu, 
and again a distinct power increase beyond $I=3.5\times 10^{21}$ \Wcmcmmu. The low energy electrons penetrate 
normally to the target surface, the energetic electrons propagate in laser beam direction.
\end{abstract}
%PACS, the Physics and Astronomy
% Classification Scheme.

%\keywords{}
%Use showkeys class option if keyword
%display desired

\maketitle

\section{Introduction}

Intense and superintense laser beam interaction with dense matter is characterized by one prominent phenomenon, that is the generation of 
superthermal high energy electrons and ions. It leads to the spontaneous question as to the effects that generate them in the absorption 
process of intense monochromatic light beams. Latest when  the kinetic temperature reaches $10^3 Z^2$ eV in the plasma, $Z$ ion charge, 
collisional absorption is ineffective and other effects of non-collisional nature have to become active in order to ensure absorption. 
The best known non-collisional candidate so far was resonance absorption at oblique laser incidence \cite{piliya,kull}. 
It consists in the direct conversion of laser light into an electron plasma wave resonantly excited at the critical electron density where the 
laser frequency $\omega$ equals the plasma frequency $\omega_p$. High intensity laser pulses in the intensity domain $I = 10^{16} - 10^{22}$ \Wcmcm\, 
with good contrast ratio are so fast rising that there 
is no time to form a preplasma in front of an irradiated solid sample that could couple to a resonantly excited plasma wave. Therefore the search begun 
for new collisionless absorption processes. The first successful proposal was the so-called ${\bf j}\times {\bf B}$ heating \cite{kruer}. The authors could 
show by particle-in-cell (PIC) simulations 
that at normal incidence the Lorentz force induces non-resonant electron oscillations at $2\omega$ normal to the target surface which lead to appreciable 
absorption, target heating and production of superthermal electrons at any density above critical. However, no attempt was made to explain how the observed 
absorption, i.e., irreversibility, comes into play. Not long after a \nobreak{remarkable} step forward was made by Brunel \cite{brunel} in understanding high-power 
collisionless absorption. He could show after introducing a few modifications that the resonance absorption concept could be adapted to steep highly 
overdense plasma profiles and significant absorption could 
be achieved under oblique incidence despite total absence of plasma resonance at $\omega = \omega_p$ ("not-so-resonant, resonant  absorption" \cite{brunel}) 
and no possibility for a plasma wave to propagate into a shallow preplasma in front of the target. Instead, now the energy imparted to the electrons is 
transported into the target and deposited there. \nobreak{Under} the assumptions of cold (i) infinitely dense plasma 
(ii) with discontinuous interface to vacuum (iii) Brunel could formulate the laser-matter interaction dynamics in the vacuum in front of the target in terms of three ordinary differential equations. 
Perhaps for this reason Brunel's mechanism of the electrons pulled out into the vacuum and then pushed back into the field-free target interior is frequently 
identified with the term "vacuum heating", an expression coined later for the particles forming a kind of thermal cloud at the vacuum-ion interface in PIC 
simulations \cite{gibbon}.

To be more specific, the reason why the term \nobreak{vacuum} heating played an ominous role in the past and partially still does presently is because, often invoked 
as the leading collisionless absorption  mechanism, it has never been defined properly. To introduce some rating in this respect it seems that two groups of 
authors can be distinguished. By the concept of vacuum heating the first group addresses the electrons in front of the sharp-edged target that circulate in 
the vacuum and do not cross the interface during one laser cycle \cite{gibbon,gibbo}. Identification of vacuum heating with Brunel's mechanism is made by 
the second group \cite{chen,getz}  
to contrast with anomalous skin layer absorption \cite{rozmus,yang,ferrante}. By the latter all motion is strictly confined to the target inside. 
Sometimes vacuum heating is interpreted as a consequence of unspecified wave breaking \cite{kato,cai}. Meanwhile it has been clarified that Brunel/vacuum 
heating prevails distinctly on skin layer absorption and that vacuum heating in the restricted sense, i.e., the contribution to absorption of the 
electrons not entering the target with the periodicity of the laser, is almost insignificant \cite{bauer}.
%\begin{minipage}{0.5\textwidth}
%\centering
\begin{figure}
\includegraphics[width=0.3\textwidth]{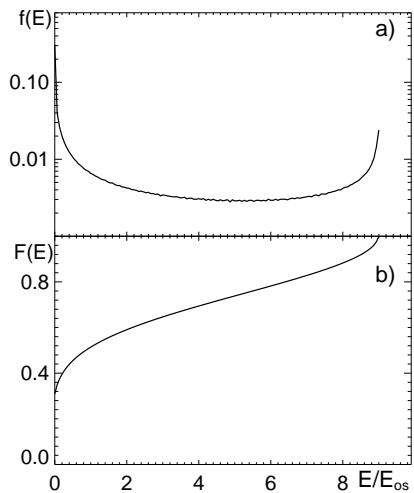}
\caption{(a) Electron spectrum $f(E)$ from Brunel's nonrelativistic model\cite{brunel}  ; 
(b) $F(E) = \int f(E')dE'$. Energy $E$ in units of mean oscillation energy in vacuum; % $E_{\rm os} = m_ev_0^2/4$; 
energy cut off is at $E = 9.1E_{\rm os}$.} 
\label{brun}
\end{figure}

Only recently a detailed analysis and discussion of Brunels's model has been given \cite{mulser}. With a view on the means of the present paper a compact 
summary of the results may be of interest. The laser field component perpendicular to the target is assumed to have the structure $E(t) = E_0\sin\omega t$. 
It generates electron jets of periodicity $\tau = 2\pi/\omega$. All of them are ejected during the first quarter period and all, except $2.2\%$, return to 
the target during the second half period $(\pi, 2\pi)$. During the remaining $3/4$ period no further electron ejection is possible as a consequence of partial 
screening by the outer layers and driver field inversion. Contrary to a common believe that all electrons in one jet are pushed back by the inverted field, 
only half of them, lifted in the interval $(0, \pi/4)$, are in phase with the driver, the other half experience a weakened driver due to screening and fall 
back to the target, attracted by the immobile ions, before the laser field has 
changed direction. This leads quite naturally to a classification into energetic and less energetic electrons. If not specified differently, throughout the paper 
we define, somehow arbitrarily, those electrons as hot whose return energy exceeds the quiver energy $E_{\rm os}$ of the free electron. Accordingly, 34 \% of 
the Brunel electrons are hot and carry 82 \% of the energy in the single jet. In contrast to PIC simulations the Brunel spectrum is non-Maxwellian with a 
pronounced maximum at $E = 9.1E_{\rm os}$ followed by a sharp cut off (see Fig.\ref{brun}). Absorption $A = I_{\rm abs}/I$ is considerably lower than measured at 
intermediate \nobreak{angles} of incidence \cite{mirela} but reaches unity at 
$86^{\circ}$ of incidence. $A$ and the absorbed energy scale like $I^{1/2}$ and $I^{3/2}$, respectively. Evidently this is the price Brunel pays for 
oversimplification. We want to stress that in Brunel's model crossing of layers among each other during the laser action is excluded, except a few front 
layers whose 
contribution to absorption is negligible. In other words, the electron flow dynamics is laminar, no wave breaking or, more appropriate in the context, no 
breaking  of flow occurs. Brunel's model offered, within limits, the first physical explanation of ${\bf j}\times {\bf B}$ heating at $2\omega$. And yet,
Brunel's model does not give a direct physical feeling for the absorption process. The numerous wrong, at least inexact interpretations of it that are still 
around are a direct indicator (example: "all electrons are pushed back by the laser field" contrasts with the true "free fall" of half of them).\\
\indent From an early one-dimensional (1D) PIC simulation the energy scaling $E_{\rm hot} \sim I^{1/2}$ has been extracted for the hot electrons \cite{wilks}; 
it has been re-'confirmed' by independent simulations \cite{baeva} and apparently by experiments \cite{malka,kenna}. However, the scaling seems to 
be  questionable for its too strong dependence on intensity; it contrasts with other experiments \cite{beg} and more sophisticated analysis \cite{haines}. 
In turn, corrections to the latter have been given recently on the basis of a relativistic kinematic model \cite{gibb}. Analogous scaling laws have been 
proposed by numerous other authors \cite{mirela,chenk,amac,kluge}. Nevertheless there is no convergence towards a definite scaling\cite{macchi}. In order to achieve further 
progress the investigation has to start from a discussion of collisionless absorption, a flow analysis of the absorbed energy into the various plasma components, 
a  definition of the hot electron component, 
and completed by analytical modeling in combination with concomitant simulations. In what follows we present our considerations on the degree of understanding 
collisionless laser beam absorption, the process of fast electron generation and their interaction with the low electron energy component in order to arrive at 
more firmly validated scaling relations  in forthcoming work. The analysis will enable us also to get insight into shortcomings of the existing models for intense 
laser-dense matter interaction.

\section{Collisionless absorption basically understood}

Let us consider the phenomenon of collisionless absorption of high-power laser beams from a more fundamental point of view. \nobreak{Under} quasi-steady state conditions 
Poynting's theorem averaged over one laser cycle reduces to
\begin{equation}
\overline{\nabla {\bf S}} = - \overline{{\bf jE}}. \label{poynting}
\end{equation}
The energy flux density is the Poynting vector ${\bf S} = \varepsilon_0 c^2 {\bf E}\times {\bf B}$; it relates to the laser intensity by 
$I = \bar{{\bf S}}$. With $n_e$ the electron density and ${\bf v}$ the mean electron flow velocity the current density is ${\bf j} = - en_e{\bf v}$. 
Equation (\ref{poynting}) describes all kinds of absorption, collisional, noncollisional, classical or quantized; in the latter case the current 
density and the electric field ${\bf E}$ are to be substituted by their operators acting on the corresponding state vector $|\psi\rangle$.  
In the intense laser field, despite the high particle densities involved, the classical picture is an excellent approximation. If the laser 
field evolves in time as ${\bf E} \sim \sin\omega t$ the current density follows as ${\bf j} \sim \cos(\omega t + \phi)$ and
\begin{equation}
\overline{\nabla {\bf S}} = - \overline{{\bf jE}} \sim - \overline{\cos(\omega t + \phi)\sin\omega t} = - \frac{1}{2}\sin\phi. \label{poyntaver}
\end{equation}
Dephasing between driver field and current determines the degree of absorption. Thus, collisionless absorption reduces to the problem of finding 
out which effects lead to a finite phase shift in ${\bf j}$. In collisional absorption it is 
the friction originating from the collisions between electrons and ions,

\begin{equation}
\overline{{\bf jE}} = \varepsilon_0\omega_p^2\frac{\nu}{\omega^2 + \nu^2}|{\bf E}|^2 > 0,  \label{friction}
\end{equation}
$\nu$ collision frequency. At $\nu = 0$ the collisional phase shift vanishes and any finite $\phi$ can only be of dynamic origin. 
Up to $I = 5 \times 10^{20} - 10^{21}$ \Wcmcm\, this dynamic origin is found in the space charge induced by $\nabla {\bf v} \neq 0$. 
The space charge generates an electrostatic field that, superposed to the laser field, determines the electron motion and leads to the 
desired finite phase shift $\phi$. This can be seen most immediately with a constant electric field ${\bf E_0}$. It yields per electron

\begin{equation}
\overline{{\bf jE}} = \frac{2\pi e^2{\bf E_0}^2}{m_e\omega} > 0.
\label{phaseshift}
\end{equation}
It is interesting to note and it can be formally shown, however it is also physically evident that 
$\overline{{\bf j}{\bf(E_{\rm Laser} + E_s})} = \overline{{\bf jE}}_{\rm Laser}$; all work is done by the driver field, 
the space charge field ${\bf E_s}$ is inert, it provides for the phase shift only. Some authors may attribute absorption to the 
Brunel like abrupt reduction of the laser wave amplitude in the skin layer. Due to this asymmetry the energy gained by an electron 
in the vacuum cannot be given back anymore to the wave when entering the evanescent region. However, for this picture to work an 
electrostatic field component is needed, too; transverse and longitudinal components cannot be isolated from each other. In the 
standard resonance absorption  at the critical density it is the space charge field of the electron plasma wave that provides for 
collisionless absorption up to $49\%$ through a phase shift $\phi \neq 0$ \cite{kull}. 
On the fundamental level of eqs.  (\ref{poyntaver}) and (\ref{phaseshift}) collisionless absorption of superintense ultrashort laser pulses 
may be classified as fully understood now for $I < 10^{21}$ \Wcmcm\, for optical wavelengths.\\
\indent All kinds of difficulties and complications arise when the degree of absorption has to be quantified. This step can only be done by 
introducing appropriate models. Numerous attempts into this direction have 
been undertaken with the intention to explain (i) the origin of the hot and the warm electron components, (ii) when and how they are created, 
during one laser period by direct resonant and nonresonant acceleration, or by stochastic processes over several laser cycles, and (iii) where is 
absorption localized, in vacuum or in the skin layer. Correspondingly, the existing absorption models may be characterized as statistic or as dynamic. 
Examples of the first class are vacuum heating in the restricted sense \cite{sentoku,rastunkov}, wave breaking\cite{kato,cai}, skin layer absorption, 
e.g.\cite{rozmus}, 
linear and nonlinear Landau damping\cite{korneev,zaretsky}. Candidates of the second class are, first of all, sharp edge absorption\cite{brunel}, 
longitudinal\cite{sofronov} and transverse\cite{pukhov} ponderomotive heating, ``zero vector potential mechanism''\cite{baeva}, relativistic 
kinematic model\cite{gibb}, anharmonic resonance\cite{mulbar}. Let our PIC simulations decide on questions (i) - (iii) and to what degree statistics is 
involved in the dynamics induced by the laser in dense targets.\\
\indent Now, after three decades of intense studies on superintense laser-matter interaction one would expect that such basic questions (i) - (iii) as 
labeled above should have found a final answer. The numerous models presented on performed experiments tell the opposite and show that 
no convergence has been reached yet. It is instructive to have a look at over 100 experimental and theoretical results on the absorption degree 
collected up to 2009 in \cite{davies}, (Fig.~1 in \cite{davies}), in the irradiance regime from $I\lambda^2 = 10^{18}$ \Wcmcmmu,\ to $10^{21}$ \Wcmcmmu. 
The absorption degrees range from 5\% to 95\% and yet at the constant irradiance of $6\times 10^{18}$ \Wcmcmmu\, absorption between 
$35\%$ and $85\%$ is reported. It drastically reflects the difficulties encountered in performing unambiguous  experiments with all essential 
parameters well defined. It is this situation that justifies still basic 1D simulations in order to learn more about which are the 
essential parameters defining the underlying physical processes. For example, a good portion of  difficulties and uncertainties arising in the 
context of hot electron scaling have their origin in different understanding of when an electron is "hot".
\section{Localization of absorption and origin of fast electrons}
We consider always linear polarized (in $y-$ direction) laser pulses impinging under $45^{\circ}$ angle of incidence onto strongly overdense fully ionized 
cold hydrogen targets with initial electron density $n_{e0}$ such that $n_{e0}\lambda^2\simeq 10^{23}$ cm$^{-3} \mu$m$^2.$ 
The interaction of the laser beam with the target is studied by 1D relativistic PIC simulations using the boost technique 
\cite{gibbon_cal}. 
If not specified differently, throughout the text we call all electrons with energies 
$E$ exceeding their mean oscillation energy $E_{\rm os}$ "hot" or "fast",
\begin{equation}
E > E_{\rm os} = m_e c^2[(1 + a^2/2)^{1/2} - 1]; \hspace {0.4cm} a = \frac{e{\bf |\hat{A}|}}{m_ec} \nonumber
\end{equation}
\begin{equation}
I \lambda^2= 1.37 \times 10^{18}a^2\, {\rm W/cm}^2\mu {\rm m}^2,\nonumber
\end{equation}
${\bf \hat{A}}$ maximum vector potential amplitude.

\subsection{Brunel model and vacuum heating}
\begin{figure}
\includegraphics[width=.5\textwidth]{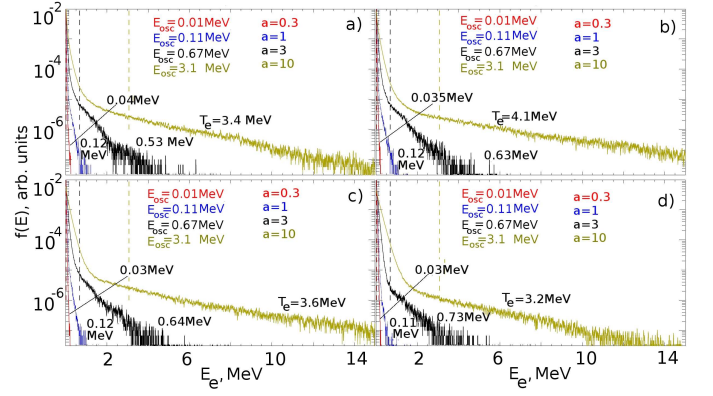}
\caption{(Color online) Electron energy spectra $\ln f(E)$ at a) 30, b) 35, c) 40 and d) 45 cycles after the 
beginning of the interaction.
A $\sin^4$ laser pulse of peak amplitude $a = 0.3, 1, 3, 10$ and full width of 50 cycles impinges under $45^{\circ}$ onto a hydrogen 
target with electron density $n_{e0}$ such that $n_{\rm{e0}}\lambda^2=9\times 10^{22}$cm$^{-3}\mu$m$^2.$  
The pulses are identical in all four frames.  
Target thickness varies from 40 to 60 $\lambda.$ Vertical dashed lines mark the mean oscillation energies. 
The hot electrons follow a Maxwellian distribution. The maximum mean energies $k_BT_e$ for $a \geq 1$ are by the factors 1.04, 1.09, 1.7 
higher than the associated $E_{\rm os}$. $k_BT_e$ increases during the evolution of laser pulse for $a = 3$, for $a = 10$ it decreases. 
Power scaling $k_BT_e \sim I^{\alpha}, \alpha \geq 0$ not  detected.}
\label{gauss}
\end{figure}

The energy spectra $f(E)$ of the hot electrons assume the typical shape of Fig.\ref{gauss}, 
here for an $I \sim \sin^4$ laser pulse of full width 50 cycles and $a = 0.3, 1, 3, 10$ after 30, 35, 40, and 50 laser cycles. 
The straight lines of $\ln f(E)$ starting from the energy $E$ between $2E_{\rm os}/3$ for $a = 10$ and from $E = E_{\rm os}$ 
for $a \leq 3$ are a clear signature of a Maxwellian type distribution function, $f(E) \sim \exp(-E/k_BT_e)$, $k_B$ 
Boltzmann constant (for example, the genuine nonrelativistic Maxwellian contains the degeneracy factor $\sqrt{E}$ to 
be subtracted from $\ln f(E)$ to yield a straight line, see related argument in Sec. V). From the slope an electron "temperature" $T_e$ is determined,  
with the significance of $k_BT_e$ the mean energy if the straight lines are extrapolated down to $E = 0$. In the Figure 
$T_e$ is indicated in units of MeV. For $a = 10$ one notices already cooling by energy transfer to the cold electrons 
and beginning plasma expansion during laser irradiation. Similar hot electron 
spectra have been reported by other authors, for example by \cite{kemp}.

%\begin{minipage}{0.47\textwidth}

\begin{figure}%[ht]
\centerline{
\includegraphics[width=0.24\textwidth]{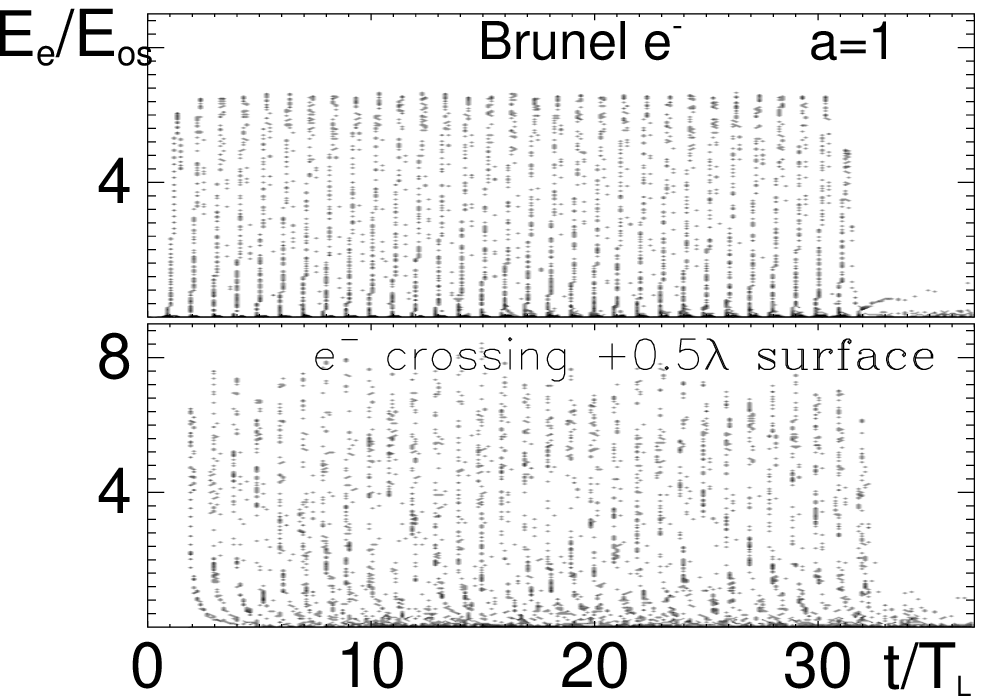}
\includegraphics[width=0.24\textwidth]{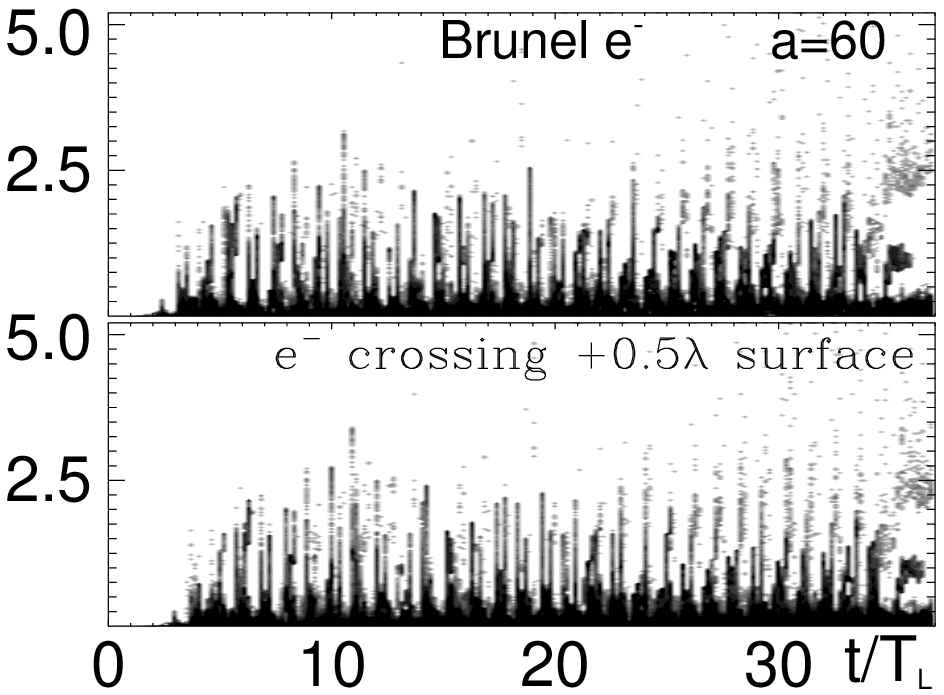}
}
\caption{Energy spectra of Brunel jets (upper pictures) and jets at depth $\lambda/2$ as function of time (units in laser cycles)
for $a = 1$ and $a = 60$. Double structure is due to ${\bf v \times B}$ acceleration. Strong reduction of $E_{\rm max}/E_{\rm os}$
with increasing $a$ is noticeable.}
\label{jets}
%\end{minipage}
\end{figure}

\begin{figure}
\centerline{
\includegraphics[width=0.23\textwidth]{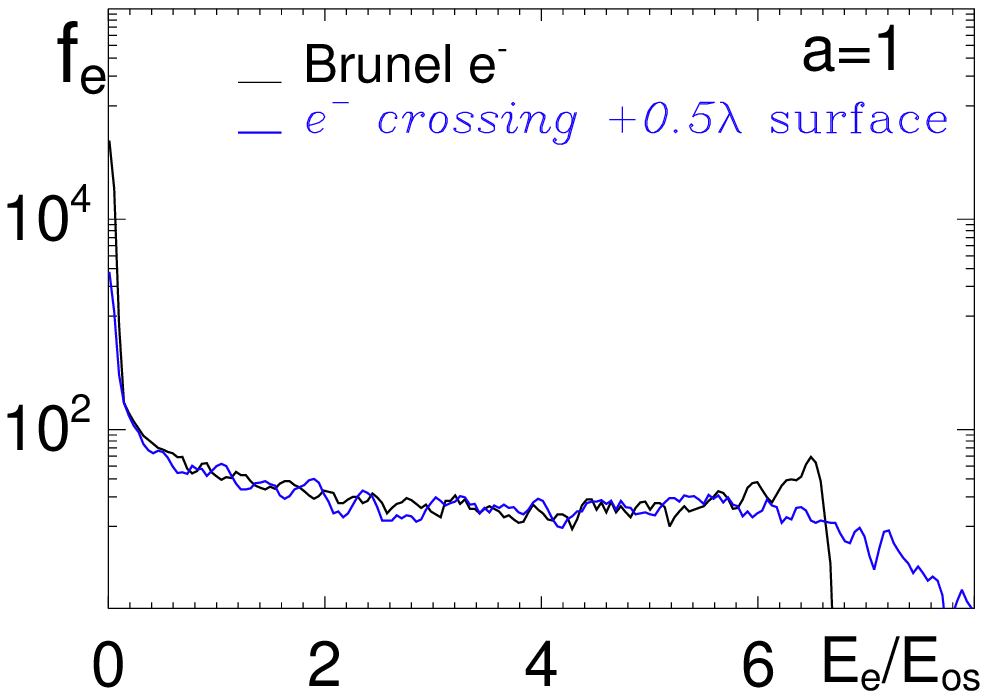}%{a=1_fd_Brunel_skin.eps}
\includegraphics[width=0.23\textwidth]{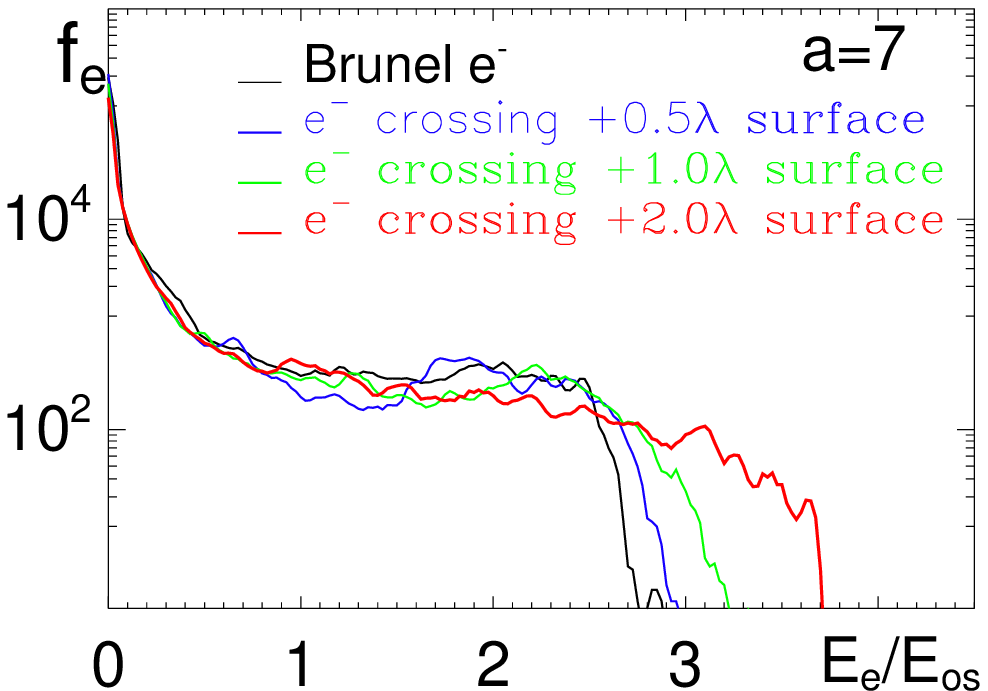}%{a=7_fd_Brunel_skin.eps}
}
\centerline{
\includegraphics[width=0.23\textwidth]{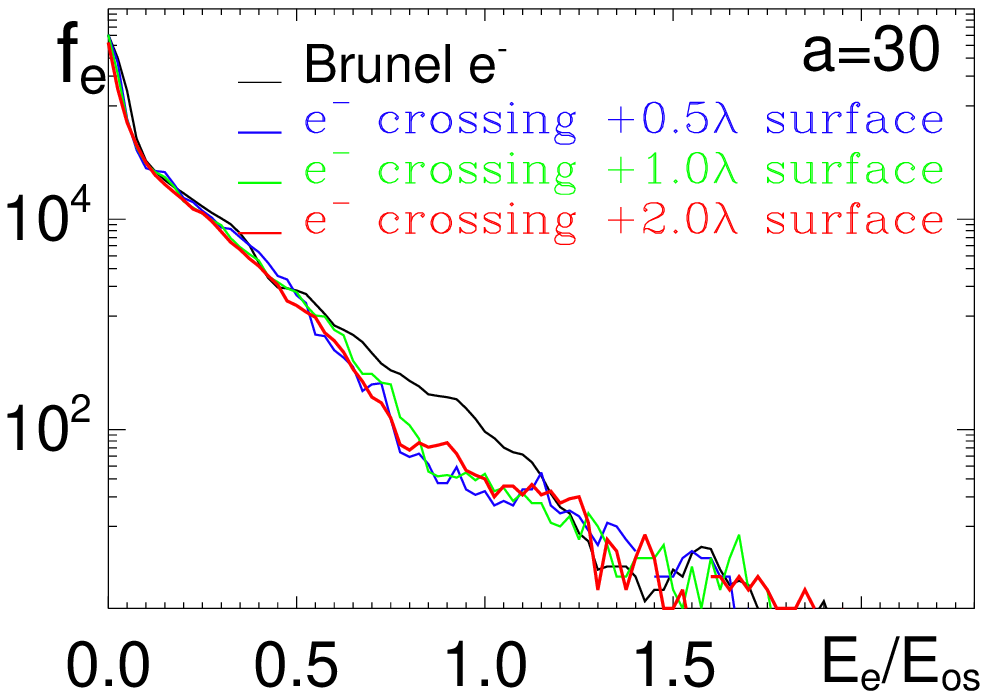}%{a=30_fd_Brunel_skin.eps}
\includegraphics[width=0.23\textwidth]{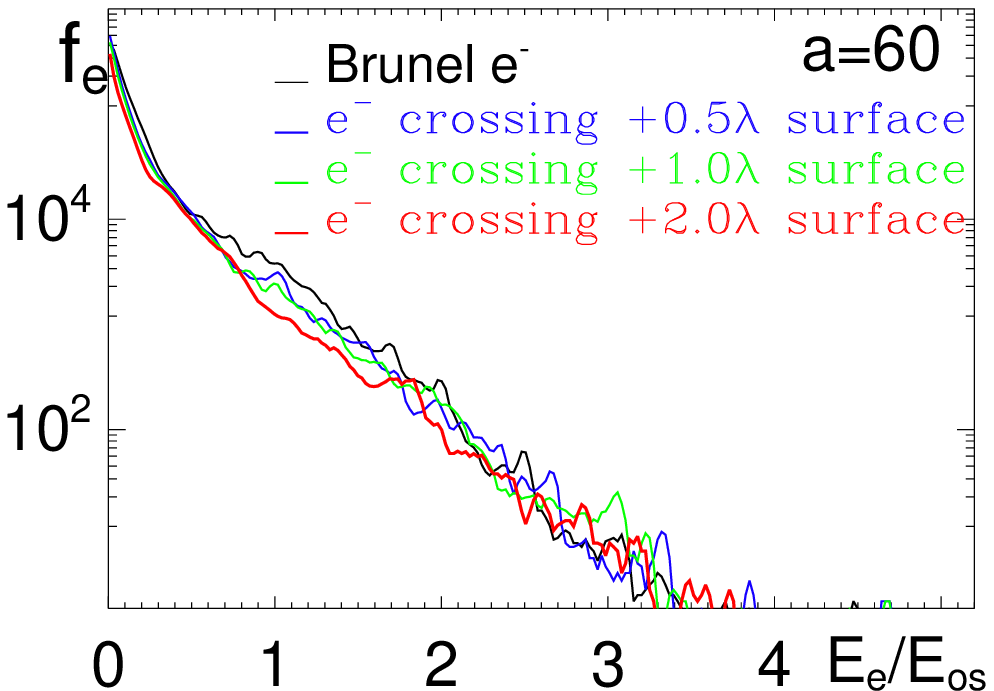}%{a=60_fd_Brunel_skin.eps}
}
\caption{Energy distributions $f(E)$ of Brunel electrons and of electrons crossing positions 
$x = 0.5\,\lambda, 1\,\lambda, 1.5\,\lambda$ and $2\,\lambda$ at intensities $ a = 1,\, 7, \,30$ and $60.$ 
The distributions are taken after 37 laser cycles when all electrons have returned to position $x = 2\,\lambda$. 
The non-Maxwellian structure of the Brunel electrons is preserved up to $a = 15$. For higher intensities there 
is almost no difference between electrons heated in the vacuum and additional heating in the skin layer region.}
 \label{spectra}
\end{figure}

\indent 
The interaction of intense laser beams with dense targets is very complex and rich of peculiar facets. 
On the other hand, consequences of basic effects, like collisionless interaction under non-harmonic resonance, 
are not clarified as they should. Here we report on phenomena which we believe will survive in 2D and 3D also.
Let us tentatively identify vacuum heating with the energy gained by all Brunel-like electrons ("Brunel electrons"). 
It is the sum of energies gained during the excursion into vacuum. This energy fraction is identified as "vacuum heating" 
and compared with the energy absorbed by all electrons. The target thickness is chosen such that no particles are reflected from 
the target backside and falsify the statistics. 
The laser beam intensity rises during one 
laser cycle to its full intensity, is subsequently held constant for 30 cycles and then sinks to zero during another full cycle. 
In Fig.\ref{jets} the energies of all Brunel particles and of all particles that have crossed the skin layer at a depth of half a 
vacuum wavelength are plotted at their crossing time for $a = 1$ and $a = 60$. The salient feature is their jet like structure 
predicted by the Brunel model. 
At the low intensity ($a = 1$) there is a clear distinction between the Brunel electrons and all electrons having undergone heating. 
The increase of the energy maxima from $6.5 E_{\rm os}$ 
to $8 E_{\rm os}$ and the more diffuse energy profiles of the jets at half wavelength in depth is a clear indication that 
some heating is localized in the skin layer, in contrast to the Brunel mechanism. At high intensity ($a = 60$) the jets assume a 
pronounced double structure due to the increased ${\bf v \times B}$ heating operating at $2\omega$, but the patterns of the two groups 
appear equally diffuse. The increase in energy of the fastest electrons is almost no longer visible (increase by $0.5 E_{\rm os}$). 
For $a > 1$ the fraction of Brunel electrons results always higher than the fraction of electrons moving inward and crossing
the boundary at $\lambda/2$ in the skin layer. The reason for the difference is to seek in the accumulation of Brunel electrons in the skin layer with 
increasing laser cycle number, randomized there and repeatedly crossing the 
target-vacuum interface before disappearing in the depth of the target.
%\begin{minipage}{0.47\textwidth}
It is instructive to  analyze the spectral distribution function $f(E)$ of the Brunel electrons and all electrons just when crossing positions 
$x = 0.5 \lambda, \,\lambda, \,1.5 \lambda$ and $2 \lambda$ for the laser intensities corresponding to $a = 1, \,7, \,30$ and 60, see Fig.\ref{spectra}. 
Surprising enough, at low intensity ($a = 1$) and, to a minor degree, also at $a = 7$ the Brunel electrons from the PIC simulations resemble much Brunel's 
analytical spectrum from Fig.\ref{brun}: The sharp cut offs and the adjacent maxima of $f(E)$ are reproduced, their positions however lie at much 
lower energies. The maximum of $f(E)$ is still clearly visible for $a = 15$ (not in the Figure), this time at $E = E_{\rm os}$, but the sharp cut off 
changes into a transition extending over $0.4\,E_{\rm os}$. The formation of a Maxwellian tail in the fast electron spectrum occurs in the skin layer 
and even deeper inside the target. From $a \simeq 20$ on no difference in the spectra from Brunel and total electrons can be observed, they are all "
thermalized". At $a = 30$ the spectra extend up to $1.4 \,E_{\rm os}$, at $a = 60$ the maximum energy is shifted to $E = 3.7 \,E_{\rm os}$. 
This is in agreement with the dependence of the fast electron number on laser intensity, see following Section \textbf{IV.A}. From Fig.\ref{jets} we 
conclude that at moderate intensities ($a \leq 15$) the skin layer contributes sensitively to the production of the most energetic electrons, 
either by laser-space charge resonance and/or by stochastic Brunel electron-plasmon interactions.

\subsection{Localization and mechanism of heating}
\begin{figure}
\centerline{
\includegraphics[width=0.25\textwidth]{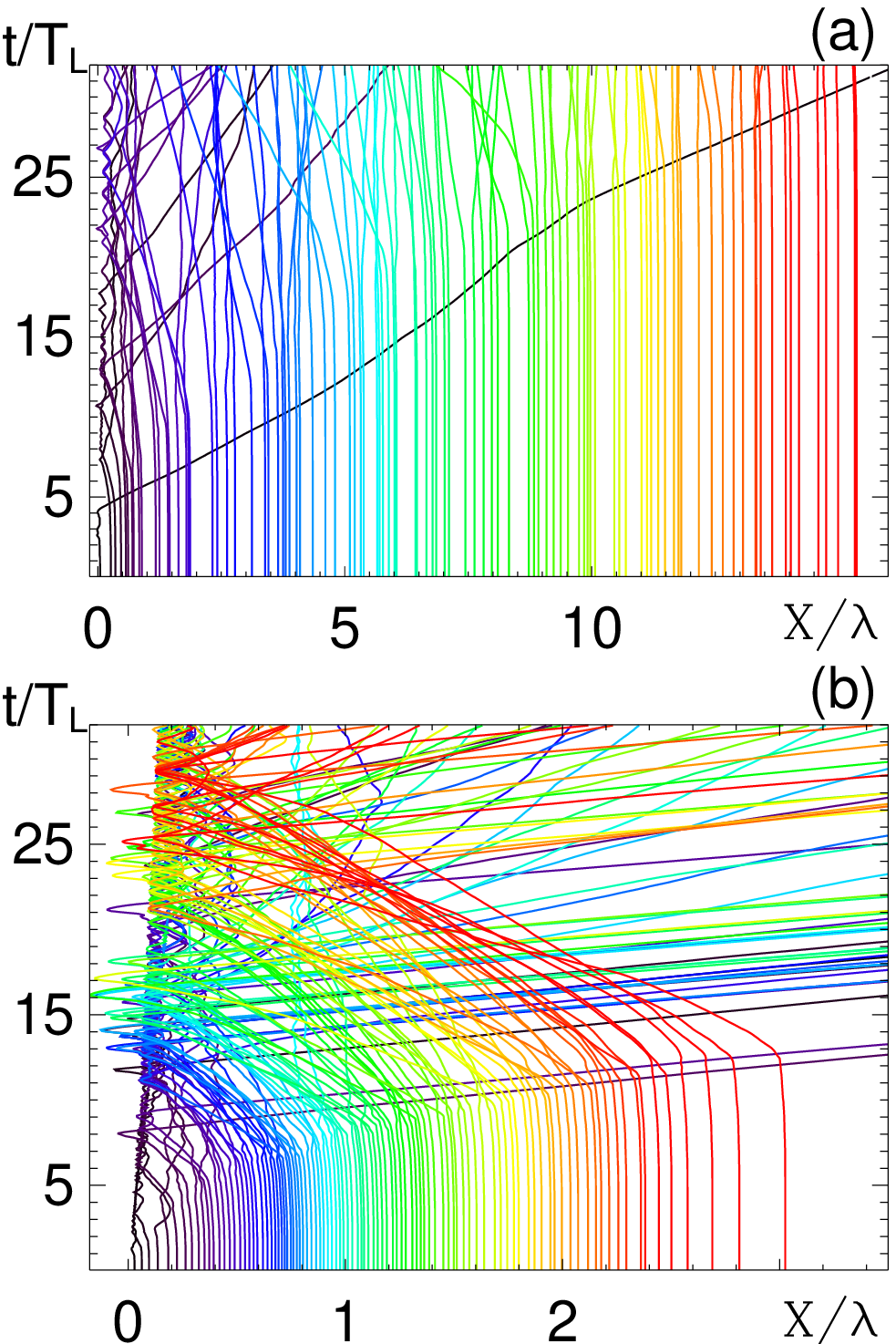}%{a=7_RED_vs_random_a=7.eps}
\includegraphics[width=0.25\textwidth]{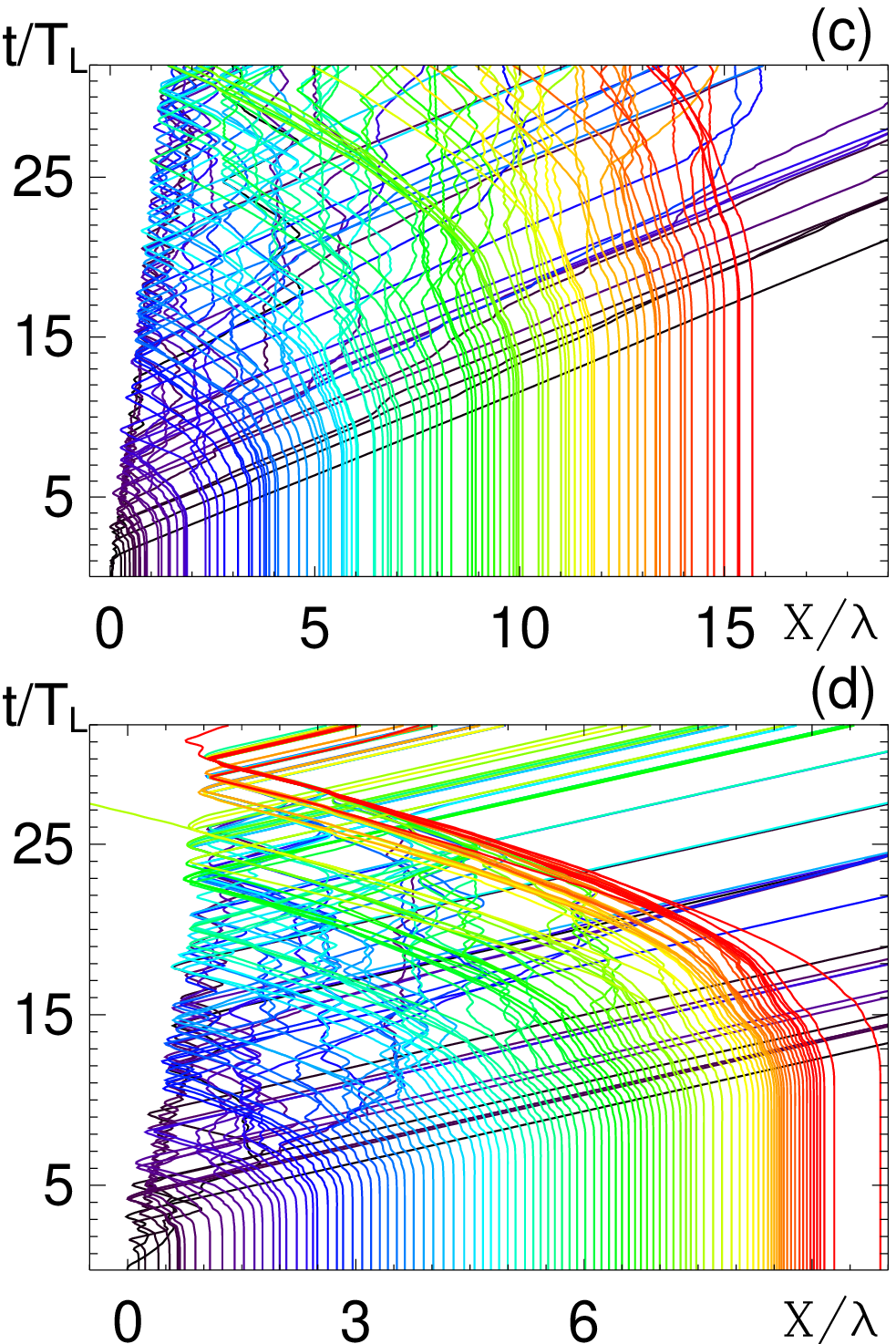}}%{a=60_RED_vs_random_a=60.eps}}
\caption{(Color online) Arbitrary selection of orbits $x_i(t), \,i = 1 - 100$, for $a = 7$ (a) and $a = 60$ (c). The lower Figures (b) and (d) show the same number of stochastically chosen orbits from the hot electrons only with $E \geq E_{\rm os}$.Their analysis shows acceleration, i.e. heating almost during one laser cycle or a fraction of it in the laser field at the target front and their strong interaction with the return current.}
\label{orbits}
\end{figure}

\begin{figure}
\includegraphics[width=0.4\textwidth]{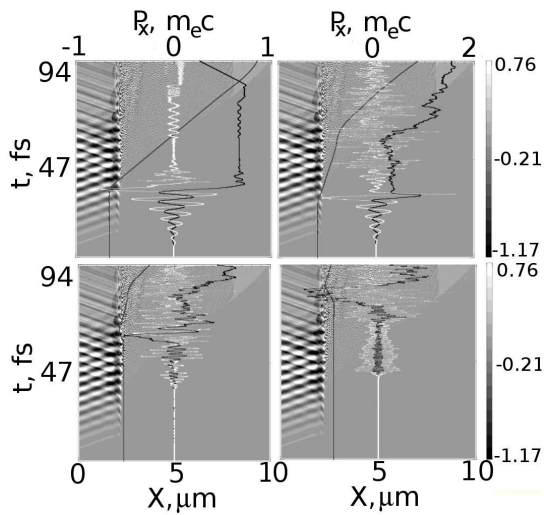}
\caption{Anharmonic electron resonance and stochastic interaction \cite{mulbar}. Regular shadow structure: 
laser field; black trajectory: orbit $x(t)$ (left) and momentum $p_x(t)$ (right), 
white traces: electromagnetic/electrostatic field at the particle's position. Primary interaction is by 
resonance between transverse and longitudinal field during a fraction of laser cycle. Particles in the last 
two pictures experience stochastic acceleration by plasmons only.}
\label{Ruhl}
\end{figure}

 Plasma density and velocity distributions as functions of time are the natural outcome in standard PIC simulations. Additional insight in the 
 heating mechanism is obtained from the orbits ${\bf x}(t)$ of randomly chosen electrons. We have analyzed numerous such computer runs each with 
 200 trajectories stochastically selected from (i) all particles heated by the laser and (ii) from the set of the hot electrons only. In Fig.\ref{orbits} 
 their time histories are depicted for the intensities $a = 7$ (left) and $a = 60$ (right). The salient features characteristic of the two groups are the 
 following:\\
(1) Heating of the energetic electrons is well localized at the vacuum-target interface and takes 
place during one laser cycle or a fraction of it. This 
excludes stochastic heating of the hot electrons. Only a low fraction of them gets the high energy in the skin layer without ever emerging into vacuum.\\
(2) Contrary to the standard assumption the "slow" return current is highly irregular as a consequence of the interaction of the jets from Fig.\ref{jets} 
with the background. It is clearly recognized in Fig.\ref{orbits} that irregular flow sets in just with the arrival of the first jets and it 
becomes the more irregular the more jets it is exposed to. The jets are accompanied by strong localized electrostatic fields that force 
electrons from the return current to reverse their direction towards the back of the target or, if they  succeed to cross the charge cloud of an incoming 
jet they are heavily accelerated towards the target front to interact further with the laser field. In short words, the laminar flow of the return current 
is heavily perturbed by the Cherenkov emission of the plasmons excited by the jets. The stochastic interaction, both, return electron acceleration and 
deceleration, has been observed in test particle models in the past \cite{bauer}.\\
(3) The plasma flow in the skin layer breaks (like "wave breaking"), i.e., the orbits cross each other, in contrast to Brunel's laminar model of infinite 
target density. \\
(4) Excursion into vacuum ("vacuum heating") of the energetic electrons decreases continuously with $a$ increasing to reach a minimum at around $a = 30$ 
and then to increase again. Owing to the significance of effects (1) - (3) for localization and understanding hot electron generation, and understanding 
collisionless laser beam absorption in general, we analyze further the acceleration process.
In Brunel's model the density of the target is assumed infinite.
Consequently all electrons start from the same position and no crossing of orbits occurs, the particle flow into vacuum and back to the target is laminar. 
Despite Brunel's oversimplification his model explains basic properties of the collisionless interaction: formation of steady state jets (Fig.\ref{jets}), 
two groups of electron energies (energetic electrons co-moving with the laser field, slow free fall electrons), 
majority of fast electrons stemming from the excursion into vacuum, dominant fraction of laser energy delivered to hot electrons. If therefore "vacuum heating" is identified with Brunel's 
mechanism it acquires a precise meaning. However there is the missing link to the physics of acceleration in this simple model. Not to forget that in 
Brunel's model all heated electrons are lifted into vacuum only during the first quarter laser period. The reality with skin layer included is different: 
The phase for lifting is stochastic, as expected from broken flow; period doubling, tripling, quadrupling, 
etc., of electron oscillations occurs in the skin layer (see Fig.\ref{orbits}); acceleration to high energies is a resonance effect. To see 
this we must concentrate once more in detail on single orbits selected statistically. In Fig.\ref{Ruhl} the time history of four %six 
particles starting 
from different depth in the target together with the electric field (white traces) they "see" during their motion is depicted for $a = 1$, i.e., 
the orbits $x(t)$ and the momenta $p_x/m_ec$ normal to the target. Resonant interaction in the first 2 pictures is clearly recognized by the abrupt 
changes in $x(t)$ and $p_x(t)$. Out of resonance the phase difference between field and momentum is $\pi/2$, see $p_x(t)$ and laser field 
(white line) in the first two %four 
pictures. The transition to resonance, i.e. field and momentum antiparallel, occurs during half a cycle or 
less in the kink of $x(t)$, seen best by zooming Fig.\ref{Ruhl}. The essential point of this resonance is its anharmonic character. In contrast to the harmonic 
oscillator in the oscillator with anharmonic potential resonance is an attractor: Given an excitation by the periodic laser above a certain threshold 
transition to resonance is unavoidable. The reason for this behavior is as follows. The harmonic potential is the only one in which the degree of excitation 
does not change its periodicity and therefore it either is driven in or out of resonance. The average stochastically perturbed space charge potential of the 
plasma is flatter than harmonic and so, depending on the excitation level its eigenfrequency changes continuously from the high level $\omega_p$  at low 
excitation down below the laser frequency $\omega$. At the crossing point resonance occurs. It has two consequences: (i) Driver, when in phase with the electron 
displacement transforms it into a runaway particle in general\cite{mulbar}; (ii) the resonant phase switch forces the electron to move against the bulk, 
the plasma flow breaks. Breaking of flow or wave breaking, respectively, often invoked as 
acceleration or absorption mechanism \cite{kato}$^,$\cite{cai} is never their origin, it is their consequence.\\ 
\indent We conclude that at $I < (5 \times 10^{20} - 10^{21})$ \Wcmcm\, the majority of energetic electrons is produced 
by resonant interaction of the laser field with the longitudinal space charge field over a fraction of one laser cycle in the vacuum 
as well as in the skin layer.
However, there is also indirect acceleration of stochastic nature of the target background, evidenced by the last two pictures in Fig.\ref{Ruhl} 
with electrons heated stochastically by the plasmons emanating from the jets. The rapidly oscillating stochastic field of the plasmons increases 
with the number of jets produced; its influence on stochastic acceleration of electrons is evident in the last two pictures of Fig.\ref{Ruhl}.

\section{Fast electrons and energy partition}
\begin{figure}
\centerline{
\includegraphics[width=0.24\textwidth]{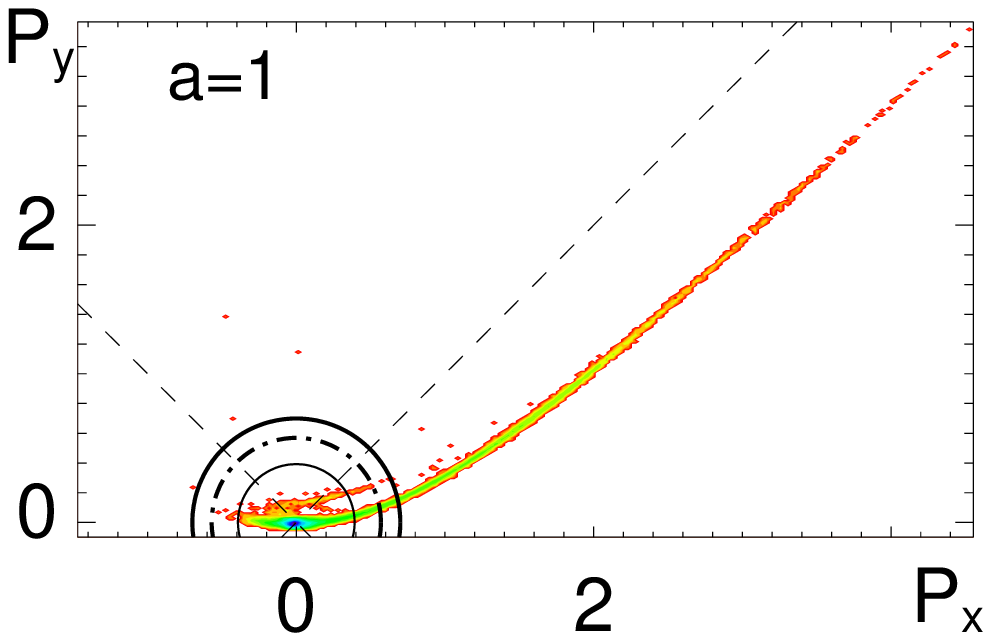}%{Px_Py_a=1.eps}
\includegraphics[width=0.24\textwidth]{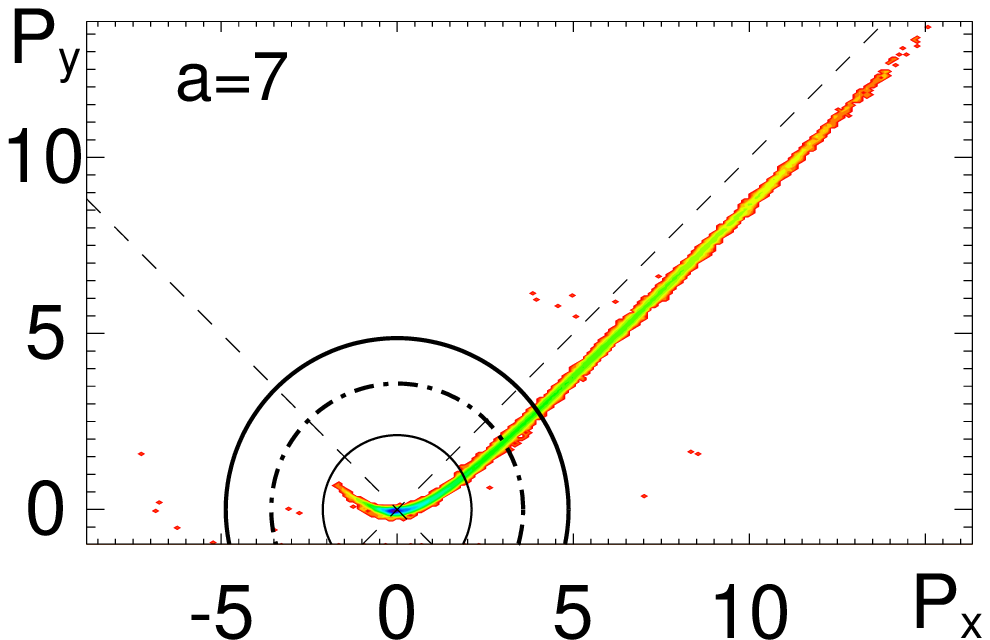}%{Px_Py_a=7.eps}
}
\centerline{
\includegraphics[width=0.24\textwidth]{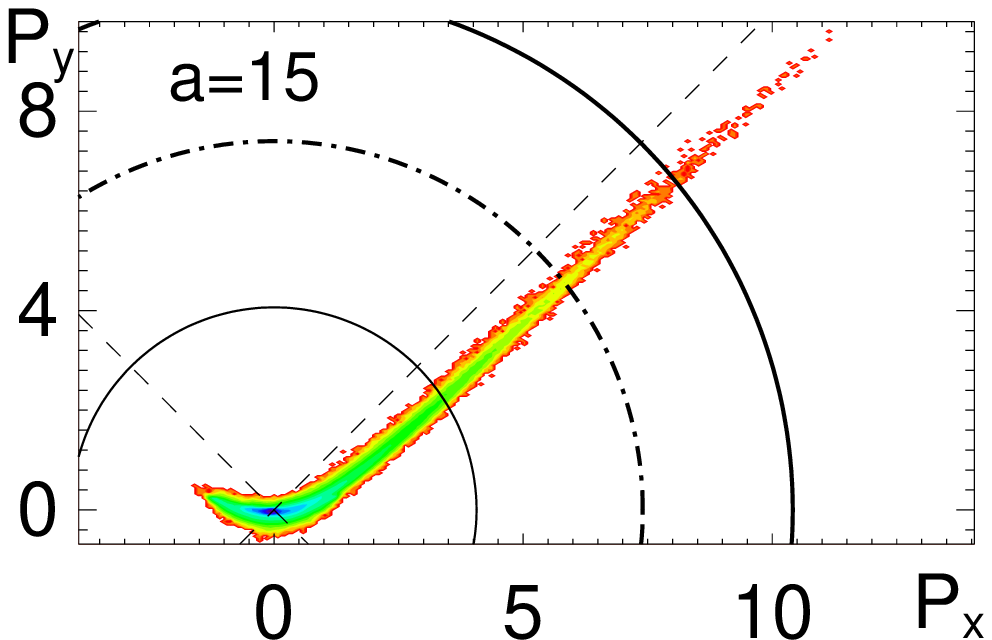}%{Px_Py_a=15.eps}
\includegraphics[width=0.24\textwidth]{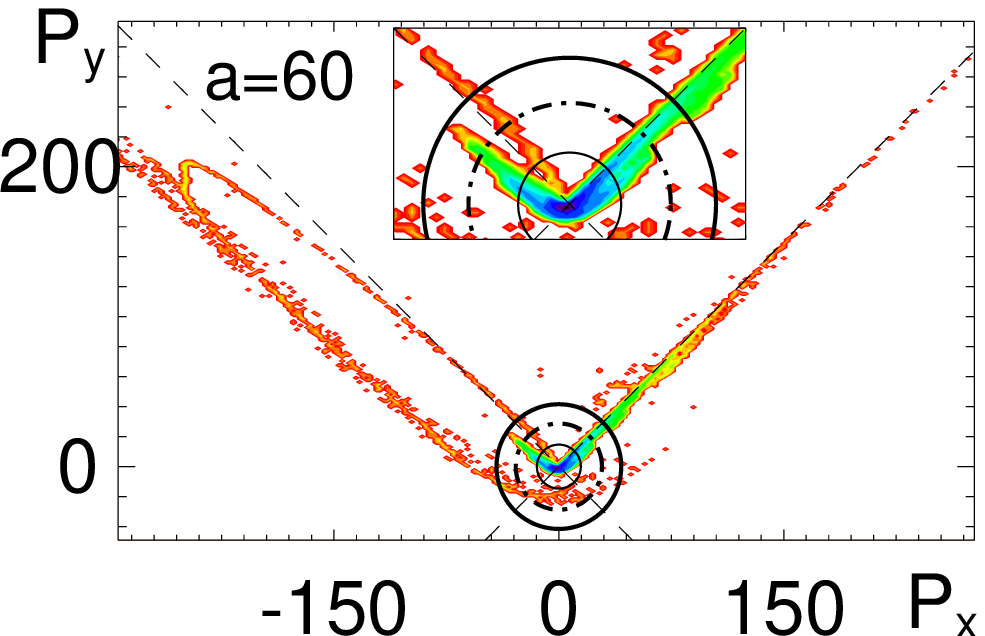}%{Px_Py_a=60.eps}
}
\centerline{
\includegraphics[width=0.25\textwidth]{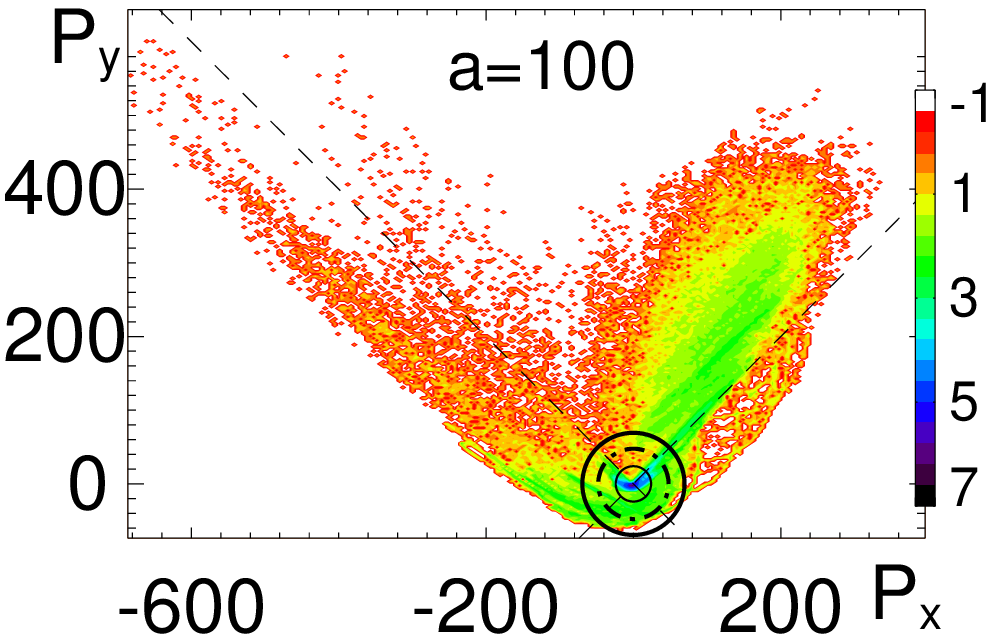}%{Px_Py_a=100.eps}
\includegraphics[width=0.25\textwidth]{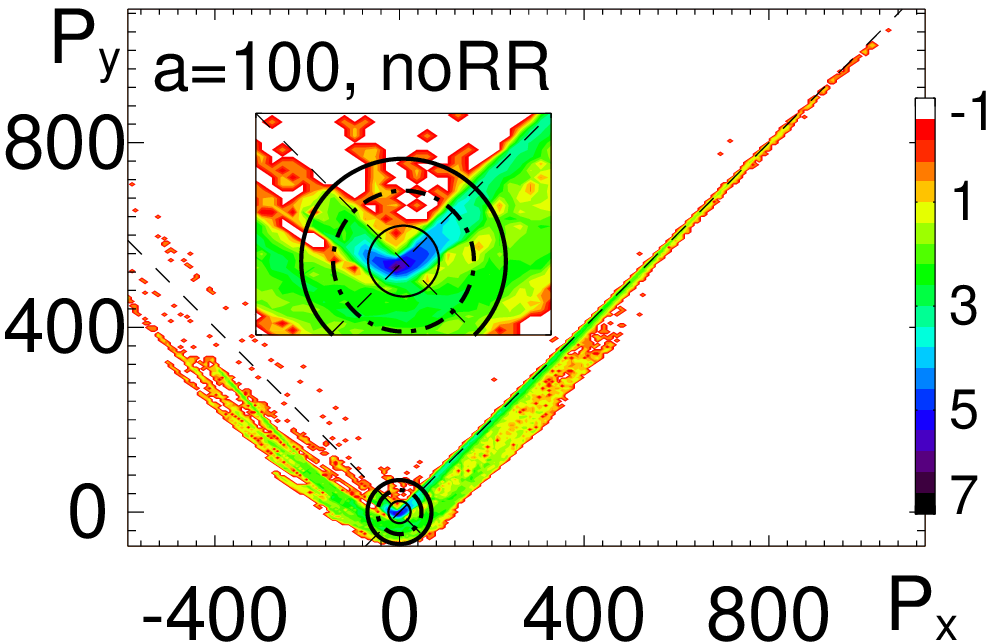}}%{Px_Py_a=100noRR.eps}}
\caption{(Color online) Direction of heated electrons: Momenta $p_y/m_e c$ vs $p_x/m_e c$ for $a = 1, 7, 15, 60$ and 100 
(left picture $a = 100$: with radiation damping, right: without) at the end of the standard laser pulse. 
Electron energies: $E<E_{\rm os}/3$ within inner black circle, $E\in[1/3, 2/3] E_{\rm os}$ within dashed 
circle, $E\in[2/3, 1] E_{\rm os}$ within bold circle, $E>E_{\rm os}$ ("hot electrons") outside. 
Circles in the last pictures are very small; therefore see the two insets). The color of the particles in this pictures indicates 
their number according to the color bar. Low energy electrons within the inner circle (majority in number) penetrate the target normally, 
energetic electrons follow the laser beam direction (dashed black lines), in $a = 60, 100$ also along the reflected laser beam.}
\label{direction}
\end{figure}

\begin{figure}
\centerline{
\includegraphics[width=0.25\textwidth]{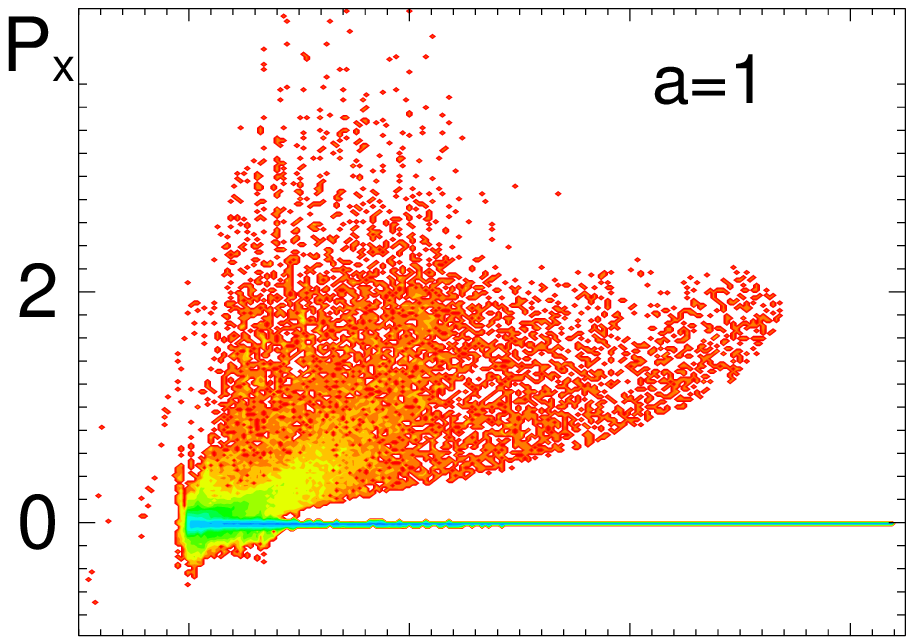}%{Px_X_a=1.eps}
\includegraphics[width=0.25\textwidth]{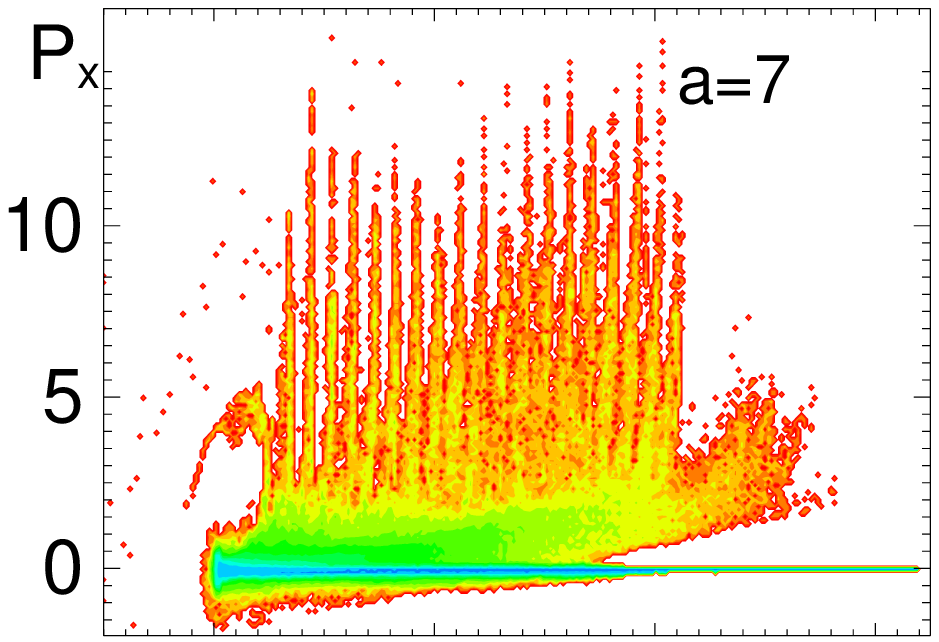}%{Px_X_a=7.eps}
}
\centerline{
\includegraphics[width=0.25\textwidth]{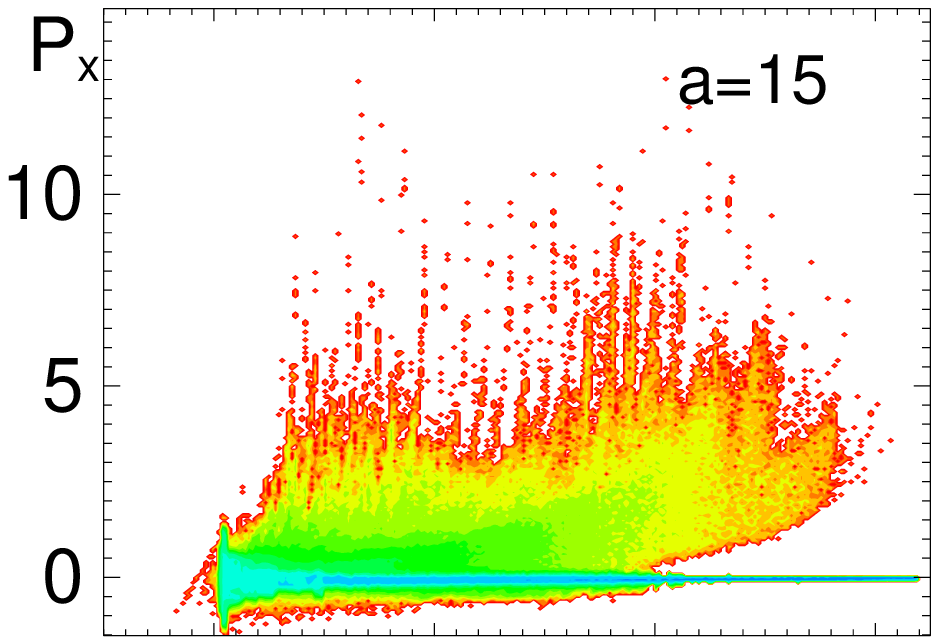}%{Px_X_a=15.eps}
\includegraphics[width=0.25\textwidth]{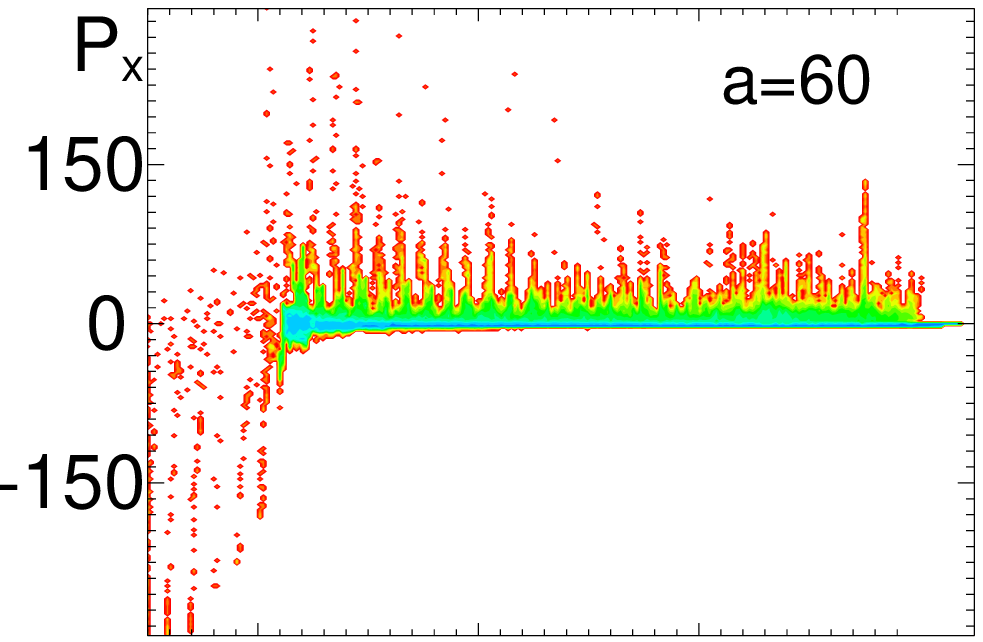}%{Px_X_a=60.eps}
}
\centerline{
\includegraphics[width=0.25\textwidth]{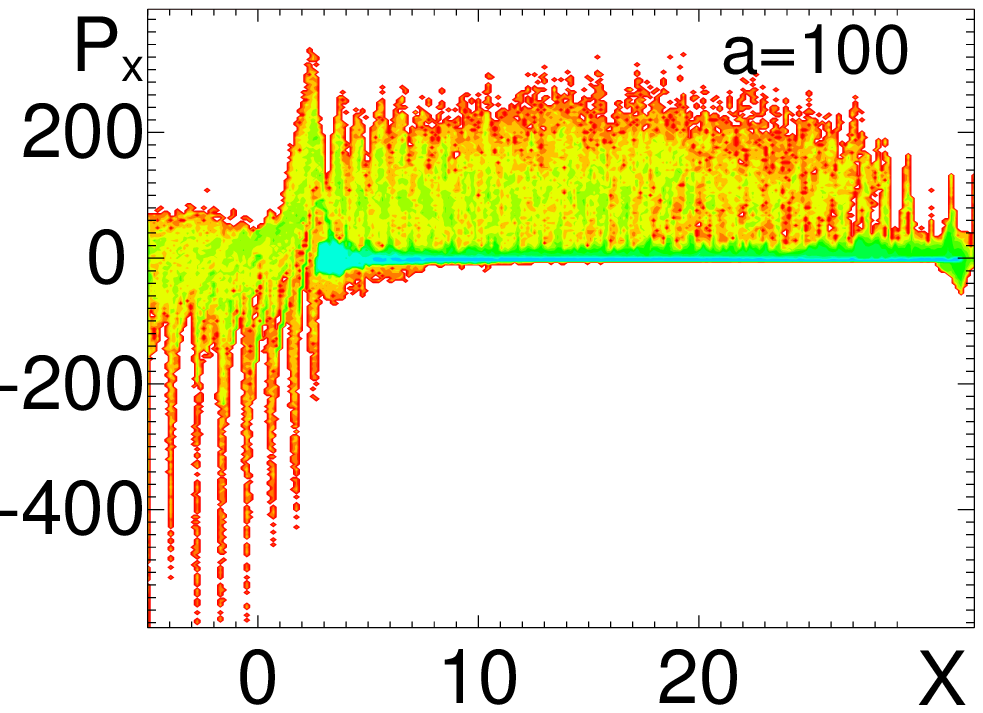}%{Px_X_a=100.eps}
\includegraphics[width=0.25\textwidth]{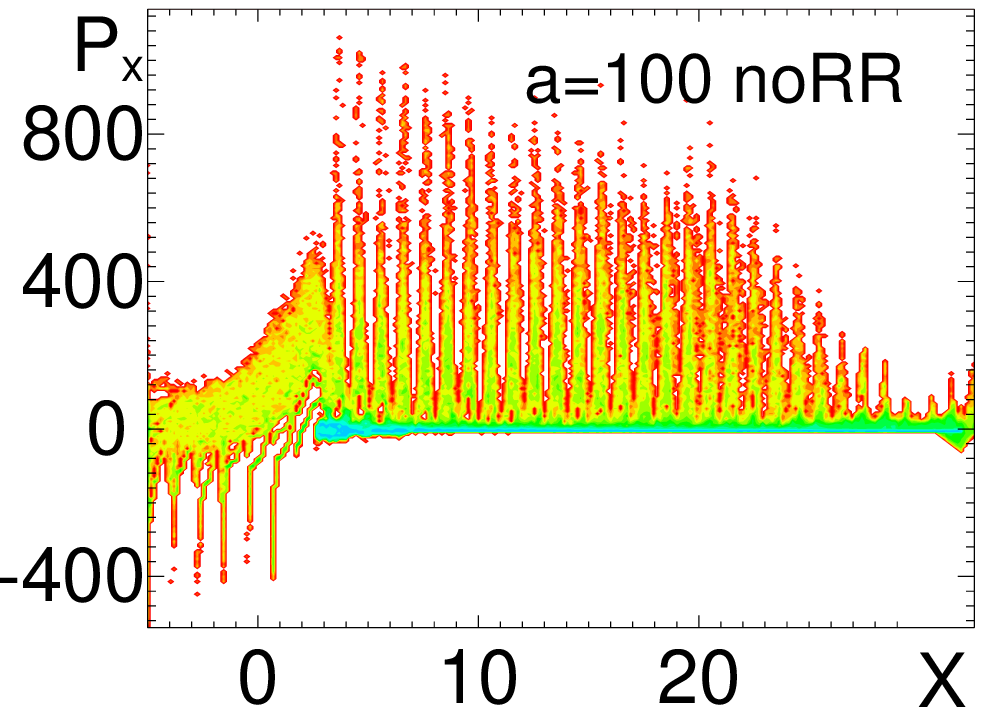}%{Px_X_a=100noRR.eps}
}
\caption{Energy distribution of heated electrons: Momentum $p_x/m_ec$ vs target normal $x$ for $a = 1, 7, 15, 60$ and 100 
(left: with radiation damping, 
right: without) at the end of the standard laser pulse. The color of the particles indicates their number according to 
the color bar of Fig.\ref{direction}. 
The spiky structure is a rough indicator of relativistic jets.}
\label{distribution}
\end{figure}

As seen in the previous section \textbf{III} there are several thermalizing mechanism: breaking of flow, skin layer noise, 
Cherenkov plasmons from jets. As a consequence one would expect that such effects dominate the low energy component of 
the electrons and that this should propagate mainly normally to the target. The more energetic an electron is the more 
it feels the Lorentz force in ${\bf v} \times {\bf B}$ direction forcing its motion into laser beam direction, $45^{\circ}$ in this paper. 
For the single free electron starting from rest in a traveling plane wave the maximum energy gain $\Delta\mathcal {E}$ and the lateral 
angular spread $\tan\alpha$ of the velocity component $v_k$ in propagation direction to the velocity component in the ${\bf E}$ 
field direction $v_E$ are\cite{mubau}

\begin{equation}
\Delta\mathcal {E} = \frac{1}{2}a^2m_ec^2,\,\,\, \tan\alpha = \left|\frac{v_E}{v_k}\right| = \frac{2}{a}, \label{velocities}
\end{equation}
thus confirming this tendency. In Fig.\ref{direction} the momenta $p_y$ parallel to the target vs $p_x$ along the target normal of the heated electrons 
are depicted for $a = 1, 7, 15, 60, 100$. The corresponding distributions of the momenta $p_x$ normal to the target over the space coordinate are shown 
in Fig.\ref{distribution}.
From the pictures it is not directly seen that the majority of slow electrons move in the direction of the target normal; 
however, as expected from (\ref{velocities}), with increasing energy the electrons follow indeed the direction of the laser beam. 
In addition, at $a = 60$ and 100 an appreciable percentage is accelerated into specular direction.
The reduction of the absolute number of hot electrons with increasing intensity, their almost vanishing at $a =15$ and their 
impressive reappearance towards $a = 100$ is particularly striking. This effect will have direct impact 
on every attempt to formulate scaling laws for the "hot electron" production. We have counted their number as a function of intensity; 
the result is reported in Table~I. Drop and increase with intensity is beyond expectation.
\begin{table}[ht]
\caption{Number of hot electrons per unit area (arbitrary units) in dependence of $a$ for $n_{e0}=100 n_c = 100 m_e\omega^2\epsilon_0/e^2.$}
\centering
\begin{tabular}{p{10mm}||p{10mm}p{10mm}p{10mm}p{10mm}p{10mm}p{10mm}}
\hline
a&1&3&7&15&30&60\\[1mm]
\hline
N$_{\mathrm{hot}}$&7819&7991&17464&147&265&19273\\[1mm]
\hline
\end{tabular}
\label{table:first}
\end{table}

The formation of spatial spikes within groups of energies and laser intensities is depicted in 
Fig.\ref{distribution}. It has to be seen as complementary to the spike distribution in time in Fig.\ref{jets}. At low laser intensity only 
the fastest electrons form jets in space as long as they are "young". As they travel further into the target they become increasingly diffuse as a 
consequence of their interaction with the Cherenkov plasmons. The electrons of varying velocity undergo 
mixing in phase space, see uniform background in Figs. \ref{jets} and \ref{distribution}, the spikes only are accompanied 
by strong elecrostatic fields. Their damping by friction is given through the collision frequency 
$\nu_{coll} = 2 (e^4 n_{e0}/\epsilon_0 m_e^2 v_e^3 \gamma) \ln\Lambda.$  With the Lorentz factor 
$\gamma = 1, v_e = c$ and $n_{e0} = 10^{23}$ cm$^{-3}$ this results in $\nu_{coll} = 5 \times 10^{10} \ln\Lambda$ s$^{-1},$ 
hence, collisional damping of spikes is unimportant.  Anomalous interaction of the laser heated electrons with the 
background has been studied recently \cite{Sherlock}. 
All electrons after having entered the distinctly relativistic regime show a neat spiky 
structure because they all fly at light speed and behave much stiffer now against their concomitant space charge field. 
The inclination of groups of spikes with respect to the normal to the abscissa at subrelativistic speeds is self-explaining. We note also that the excursion of 
the slow electrons into vacuum ($p_x$ negative) reduces with increasing $a$.

\subsection{Partition of the absorbed energy}
Sometimes it is claimed (at least in the past) that all electrons are "hot" in intense laser-solid target interaction. This rises the question on the 
percentage of the hot electrons with respect to number, to average energy, or to average flux density. Here, we must stress that a percentage in particle 
number cannot be given, neither in the experiment nor in the simulation for the simple reason that the fraction depends very sensitively on the total number 
of particles involved: Where to put the lower threshold? Should the shock heated portion of the target be counted also, or is it reasonable to restrict counting 
on those electrons that have "seen" the laser at least once? However, the situation is totally different with respect to energy 
fractions because no ambiguity arises on that. 
In Table II we present such an absorbed energy partition as a function of laser intensity (parameter $a \sim I^{1/2}$) for two overdense targets, 
$n_{\rm e0} = 100 n_c$ and $n_{\rm e0} = 200 n_c$ (for $a = 15$ also $n_{\rm e0} 
= 400 n_c$): overall fraction of absorption $I_{abs}$; percentage of energy which is found in the electrons; total fraction of energy absorbed by the hot electrons, $E \geq E_{\rm os}$ and by hot + medium hot electrons  of $E >  E_{\rm os}/2$ ("warm $e^-$"); energy fraction transmitted to the ions; energy fraction found in the electrostatic space charge field ("fields"). 
A first view on the Table tells that the main
absorption is accomplished by the energetic electrons (see 5$^{th}$ and 6$^{th}$ column). 
The absorption by the ions (protons in the Table) remains modest for $a \leq 15$, however, Cherenkov plasmons ("fields") assume a 
non negligible portion of laser energy, more than we predicted.
\begin{table}
\caption{Partition of the incident laser energy: fraction of absorbed intensity $I_{abs}$ transmitted to the electrons, 
the hot and warm electrons, the ions and the plasmons ("fields") at the end of the standard laser pulse.}
\centering
\begin{tabular}{c|c|c|ccccc}
\hline
\multirow {2}*{$a_0$}&\multirow {2}*{$n_{e0}/n_c$}& \multirow {2}*{A} & \multicolumn{5}{c}{Energy partition}\\
%\hline
\cmidrule(l){4-8}
%\cline{4-9}
&&&all $e^-$&hot $e^-$&warm $e^-$&ions&fields\\
\hline
\multirow {2}*{0.3}&100&0.377&0.25&0.213&0.217&0.009&0.118\\
                   &200&0.313&0.24&0.161&0.167&0.007&0.066\\
\hline
\multirow {1}*{0.5}&100&0.43&0.28&0.228&0.233&0.01&0.138\\
                   %&200&&&&&&\\
\hline
\multirow {2}*{1}&100&0.358&0.238&0.2&0.211&0.008&0.112\\
                 &200&0.354&0.24&0.199&0.206&0.0077&0.106\\
\hline
\multirow {2}*{3}&100&0.18&0.122&0.08&0.092&0.003&0.055\\
                 &200&0.18&0.123&0.08&0.092&0.003&0.054\\
\hline
\multirow {2}*{5}&100&0.2&0.136&0.088&0.102&0.0033&0.061\\
                 &200&0.2&0.136&0.089&0.103&0.0026&0.061\\
\hline
\multirow {2}*{7}&100&0.19&0.131&0.073&0.089&0.0033&0.056\\
                 &200&0.19&0.132&0.076&0.093&0.0025&0.055\\
\hline
\multirow {3}*{15}&100&0.067&0.036&0.0002&0.003&0.01&0.021\\
                  &200&0.051&0.028&0.0005&0.003&0.007&0.016\\
                  &400&0.064&0.039&0.0022&0.008&0.0057&0.019\\
\hline
\multirow {2}*{30}&100&0.105&0.0525&0.0002&0.0018&0.0206&0.032\\
                  &200&0.045&0.0168&0&0.00012&0.0168&0.011\\
\hline
\multirow {2}*{60}&100&0.23&0.126&0.01&0.027&0.033&0.071\\
                  &200&0.092&0.031&0.00006&0.0003&0.034&0.027\\
\hline
\end{tabular}
\label{table:second}
\end{table}

The increase in ion energy beyond $a = 15$ is due
to the deeper penetration of the laser as a consequence of the recession of the electrons by the radiation pressure and hence increased energy 
coupling to the ions.

\begin{figure}
\centering
\includegraphics[width=0.4\textwidth]{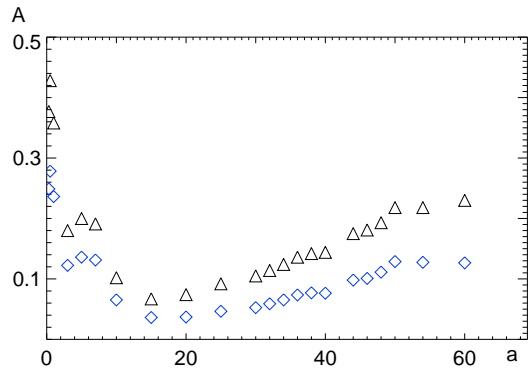}
\caption{(Color online) Total absorption (triangles) of a 30 cycles standard laser pulse (see text) and the absorption by 
electrons (blue diamonds) is given as a function of $a$. The reduction is due to oscillation inhibition by the induced 
electrostatic field, its rise beyond  is mainly a consequence of entrainment ("runaway electrons").\\}
\label{absorption}
\end{figure}

The overall absorption drops continuously with increasing laser intensity. In contrast to the runaway energy $\mathcal {E}$ in (\ref{velocities}) 
the free quiver energy at fixed oscillation center is $E_{\rm os} = mc^2[\sqrt{1 + a^2/2} - 1] \sim I^{1/2}$. However this reduction is counterbalanced 
by the relativistic increase of the critical density, $n_{cr} \sim \gamma n_c$. As the speed of the moderately hot electrons approaches $c$ the absorption 
into energetic electrons, which is the major portion, should not change; the drop  must have a different, nonrelativistic origin. Although the scaling of 
$E_{\rm os}$ and $n_{cr}$ may be oversimplified (see \cite{weng} for $n_{cr}$ scaling) it is correct in its tendency. 
Our current explanation attributes the very pronounced reduction of absorption to the limiting effect of the electrostatic field on the oscillation amplitude of the single electron: With increasing 
intensity $I$ the electrons are pushed more and more inward by the radiation pressure. The electron oscillating in the neighborhood of the vacuum-ion 
interface oscillates in a narrow anharmonic potential the half width of which towards the target interior is a small fraction of the wavelength 
("profile steepening"). A similar reduction of absorption has been reported for normal incidence, with 
an explanation that agrees qualitatively with ours \cite{sanz}. Latest beyond $I \simeq 10^{21}$ \Wcmcm\, absorption by the fastest electrons increases again. They are runaway electrons. 
The electrostatic potential is strong but finite. The phase at which the electrons enter the laser beam is stochastic. Within them there will some of 
them happen to be in resonance with the field and subject to the Doppler shift
\begin{equation}
\omega' = \gamma(\omega - {\bf kv}) \label{doppler},
\end{equation}
with $\gamma$ Lorentz factor, ${\bf k}$ wave vector. If such an electron is moving inward from the vacuum it sees the incident wave at a Doppler 
downshifted low frequency and is accelerated over a longer distance whereas the reflected  wave is seen at a highly upshifted frequency and 
represents merely a high frequency disturbance. For an electron moving outward towards the vacuum the accelerating field is that of the reflected wave. 
The proof of this acceleration mechanism is based on the study of single particle motions, and is directly confirmed by the appearance of energetic electrons 
flowing into vacuum in the reflected wave direction in the picture for $a = 60$ and $a = 100$ of Fig.\ref{direction}. To give a numerical example of electron 
displacement lengthening $\Delta x/\lambda$ in a plane TN:SA laser 
wave ($\lambda = 800$ nm) during a forth cycle $\Delta \varphi = \pi /2$: $\Delta x/\lambda = 3$ at $I = 10^{21}$ \Wcmcm\, and
$\Delta x/\lambda = 30$ at $10^{22}$ \Wcmcm. For comparison, at $I = 10^{18}$ \Wcmcm\, this shift is 0.03 only. Beyond $I = 10^{22}$ \Wcmcm\,  
radiation reaction on the electron motion has to be taken into account 
%and only for the counterpropagating phase of the motion 
\cite{naum,lis,puk}. A summary of absorption into all plasma channels (electrons, ions, plasmons) its fraction into electrons, 
the decrease of absorption towards a pronounced minimum close to zero at $a \simeq 15 - 20$ and its rise beyond is presented in Fig.\ref{absorption}.

\section{On scaling laws of the "hot electrons"}
From intensity scaling the experimentalist and theoretician expect analytical formulas of the shape of the electron spectrum as a 
function of the laser intensity. 
As such a goal seems to be beyond reach at present the high power laser community has limited its focus on the energetic electrons. There, the generation 
of a Maxwellian tail is one of the characteristics of high power interaction. It is also the most interesting part of the spectrum because, as seen from 
Table~II it contains the main part of the absorbed energy and, last but not least, it is relevant to applications for collective ion acceleration, radiation 
sources, medical applications, and others. It is aimed at how the number of energetic electrons, the degree of absorption and the mean energy scale with 
intensity. On the basis of present knowledge scaling of the first two quantities is not feasible. Regarding the mean energy, or the hot temperature 
$k_BT_{\rm hot}$, respectively, despite the frequent attempts in experiment and theory no convergence 
has been achieved so far at all. In the light of our foregoing analysis there is not much surprise about.\\ 
%\begin{minipage}{0.48\textwidth}
%\end{minipage}
The frequently invoked ponderomotive scaling ("Wilks' scaling")\cite{wilks} of $T_{\rm hot} \sim I^{1/2}$ is based on the idea that each laser cycle 
energetic electrons with energy average in the range of about $E_{\rm os} =  m_ec^2(\sqrt{1 + a^2/2} - 1)$ are generated. Subsequently this scaling has 
been recognized as too strong and, in first place guided by experiments \cite{beg}, has been replaced by the milder power law \cite{wilk} 
$T_{\rm hot} \sim I^{0.34 - 0.4}$ in the intensity range $10^{18} - 10^{21}$ \Wcmcm. It is intended as to be applied to the unidirectional Maxwellian 
electrons reaching the analyzer at time $t = \infty$. The search for the right mean energy scaling is equivalent to the search for the process of absorption. 
In the absence of the stochastic element inherent in collisions it is important to understand whether a Maxwellian tail is one of the signatures of the 
interaction and, if it is, why.

To arrive at a Maxwellian distribution in an ensemble it is sufficient to know that, given a certain amount of particles $n_{\rm hot}$ containing the 
amount of energy $\mathcal {E}_{\rm hot}$, all possible states in the relevant phase space are equally likely and that the Hamiltonian is given by the sum
\begin{equation}
H = \sum_{i = 1}^{n_{\rm hot}} \sqrt{m_e^2c^4 + c^2{\bf p}_i^2};\,\,\,\,|H| = \mathcal {E}_{\rm hot} \label{ham}
\end{equation}
which expresses the property that the single electrons are uncorrelated. If the relevant phase space is $\{({\bf p}, {\bf q})\}$, as for example in statistical 
thermodynamics, the resulting distribution is the Maxwellian momentum distribution $f(E) \sim \sqrt{E}\exp(-E/k_BT_{\rm hot})$, in disagreement with 
Fig.\ref{gauss}. However, as outlined in the foregoing chapters, fast electron generation is by anharmonic resonance. 
This has the important properties: (i) resonance is an attractor for all electrons above a certain oscillation energy, in contrast to harmonic resonance; 
the always present crossing of trajectories is a clear indicator of it. (ii) All $n_{\rm hot}$ electrons have the same chance to resonate anharmonically 
regardless of their phase with respect to the laser driver. This makes it very likely  that the relevant phase space is the energy acquired at resonance 
rather than the momentum. Then from (i) and (ii) follows that collisionless absorption is accompanied  by a Maxwellian tail of energetic 
electrons (consequence of anharmonic attractor) and the spectrum scales like $f(E) \sim \exp(-E/k_BT_{\rm hot}),$   
without a degeneracy factor $\sqrt{E}.$ This is what we also deduce from 
Fig.\ref{gauss}.
%\begin{minipage}{0.48\textwidth}
\begin{figure}
\centering
\includegraphics[width=0.4\textwidth]{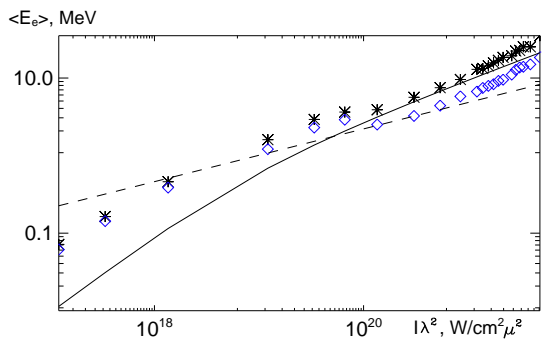}
\caption{(Color online) Hot and warm electron energy scaling $\langle E_e\rangle$  with laser intensity $I = 10^{17} -  5\times 10^{21}$ \Wcmcm, 
standard pulse. Stars: $E_e \geq E_{\rm os}$ ("hot electrons"); diamonds: $E_e \geq 0.5 E_{\rm os}$ ("warm electrons"). 
In contrast to Fig.\ref{gauss} here $\langle E_e\rangle $ is the average taken over the single energies $E_e$. 
Solid line: scaling after \cite{livrep}, $k_B T_{\rm hot} = m_ec^2[\sqrt{1 + (I\lambda^2)/(2.74\times 10^{18})} - 1]$. 
Dashed line: scaling  after \cite{beg}, $k_B T_{\rm hot} = m_ec^2 [I\lambda^2/(1.37\times 10^{18})]^{0.34}$.\\}
\label{scal2}
\end{figure}

\indent Let us first examine  Fig.\ref{gauss} for pulses $I \sim \sin^4$ in the intensity range $10^{18} - 10^{20}$ \Wcmcm. 
The uncertainties on the mean energy (slope of log scale) in pictures a) (30 cycles) and c) (40 cycles) are considerable, 
nevertheless we can conclude with certainty that neither Wilks' original ponderomotive scaling\cite{wilks} ($I^{0.5}$) nor its improved version are met 
to some extent. They are far too weak. However, the analysis shows that the assumption $k_BT_{\rm hot} = \kappa \times E_{\rm os}$ with the 
constant $\kappa$ not far from unity works. This means that at these relatively low intensities (from $a \simeq 3$ to $a = 10$) the scaling is\\
\begin{equation}
k_BT_{\rm hot} \sim \sqrt{1 + a^2/2} - 1 \label{scal1}
\end{equation}
in agreement with \cite{livrep}. From $a = 12$ on $E_{\rm os}$ is well approximated by $m_ec^2 a \sim I^{1/2}$. 
In Fig.\ref{scal2} we extended the search for scaling from $a = 10$ up to $a = 60$ the latter is already in the runaway 
regime of absorption. Satisfactory agreement with \cite{livrep} is obtained for $I$ from  $10^{20}$ to $10^{21}$ \Wcmcm. 
Beyond the change in the absorption mechanism and the stiffer coupling to the ions is noticeable in the increase of slope relative to \cite{livrep}.

\section{Summary and conclusion}
The focus of the present paper is on the physics of collisionless absorption of intense laser beams in dense 
targets in the intensity domain $I = 10^{18} - 10^{22}$ \Wcmcm for optical wavelengths, on the variation of the spectral composition of the 
energetic electrons with intensity and on their scaling with the latter. Most remarkable results are the Brunel like spectral hot electron distribution at the 
relativistic threshold, the minimum of absorption at $a \cong 15 - 30$, the drastic reduction of the number of hot electrons in this domain and their 
reappearance beyond, the strong coupling with the return current beyond expectation, and a strong hot electron scaling in $a \cong 1 - 10$, a scaling 
in vague accordance with current published estimates in $a \cong 10 - 50$ and a strong increase beyond.

On a fundamental level understanding collisionless absorption is equivalent to the search for the non-orthogonality of induced current density to the laser field. 
The answer is found in the interplay of the laser field with the space charge field induced by it. The idealized model of Brunel works already on this basis. 
It is capable of explaining important effects at the relativistic intensity threshold and below, like the generation of two groups of electrons, a hot and a 
cold component. The non-Maxwellian spectrum predicted by the model is found in our simulations at the relativistic threshold and below; with increasing $a$ it 
is washed out. By following test orbits we are able to localize absorption at the vacuum-target interface and skin layer for all intensities below the radiation 
reaction limit at 
$I \cong 10^{22}$ \Wcmcm\, in linear polarization, in agreement with Brunel for non relativistic intensities. If therefore the ominous "vacuum heating" 
is invoked as responsible for absorption this is correct 
if it is identified with Brunel's mechanism. What it does not explain, and Brunel does not either, is the underlying physics, 
i.e. the phase shift and, in concomitance, orbits crossing.

An explanation in terms of physics has to show that (i) such a breaking of flow is not by accidence and (ii) a hot 
Maxwellian tail in the spectrum is a natural outcome from strong drivers. Anharmonic resonance is currently the best model explaining both aspects. 
It rules stochastic heating and "weave breaking" out automatically. Anharmonic resonance constitutes an attractor (fix point). This kind of 
resonance always happens in presence of a sufficiently strong driver at any laser frequency and any target density, in contrast to harmonic resonance 
which is bound to $\omega  = \omega_p$. When crossing resonance the momentum of an electron undergoes a phase shift by $\pi$ or a fraction of it with 
respect to the bulk of the plasma. Wave breaking, here  more appropriately called breaking of flow owing to profile steepening on lengths of a small 
fraction of a laser wavelength, is a consequence of absorption and energetic electron generation, not its origin.

From $I \cong 5 \times 10^{21}$ \Wcmcm\, anharmonic resonance is strengthened by the generation of runaway electrons due to trapping in both, the incident 
and the reflected laser wave. The reduction up to nearly disappearance of hot electrons $E \geq E_{\rm os}$ is attributed to oscillation inhibition by the 
ponderomotive space charge field. In the whole intensity domain considered the major fraction of laser energy is deposited in the hot and moderately hot 
electrons. The next significant portion goes into Cherenkov plasmons excited by the periodic plasma jets. From the analysis of the test trajectories their 
coupling to the neutralizing return current is apparent. We consider it as an important aspect when modeling anomalous transport of heat and fast electrons 
in compressed matter.

Finally, we reexamined the hot electron scaling in 1D, perhaps the most controversially discussed subject in the pertinent literature.  Acceptable 
coincidence with the leading approximations is only found in the intensity range $10^{20} - 10^{21}$ \. The deviations below $a \cong 10$ are to be 
attributed mainly to the imprecise proportionality between $E_{\rm os}$ and $I \sim a^2$. In the runaway absorption regime the governing scaling law is 
still to be discovered. The main reason for the current misunderstandings and disagreements have to be attributed to the poor knowledge of the electron 
energy spectrum $f(E)$. What is missing most at present in the experiment and in the theory is a clear definition of what means "hot" and "cold" electrons. 
In order not to fall into this deficiency we define electrons as "hot" and "moderately hot" if their energy is higher than $E_{\rm os}$ 
and $(0.5 - 2/3)E_{\rm os}$. It is approximately the range of the Maxwellian tail. The commonly used terminology "Maxwellian" is misleading because it refers to the electron velocities ${\bf v}$ or momenta with a distribution 
law $df(E) = v^2 \exp(-E/k_BT_{\rm hot})dv \sim \sqrt{E}\exp(-E/k_BT_{\rm hot})dE$. This differs from our findings (and, implicitly, others) of a Boltzmann 
distribution $df(E) = \exp(-E/k_BT_{\rm hot})dE$ for the relevant restricted phase space of total energies $\mathcal {E} = \sum E_i$.
\begin{acknowledgments}
This work was in part supported by the DFG within the SFB 652. % and by the Program No.43 of the Fundamental Researches of the 
%RAS Presidium on the Strategical Areas of Science Development "Fundamental Problems of Mathematical Modeling". 
 PIC simulations were performed using the computing 
resources granted by the John von Neumann-Institut for Computing (Research Center J\"{u}lich) under the project HRO01. 
We gratefully thank Dr. Andrea Macchi (University of Pisa, Italy) for providing us his 1D PIC code and 
Prof. Su-Ming Weng (Jiao Tong University, Shanghai) for his contributions of ideas and suggestions. 
The continuous support of this work by Prof. Dieter Bauer (University of Rostock, Germany) is gratefully acknowledged.
\end{acknowledgments}

\end{document}